\documentclass[10pt,letterpaper]{article}
\usepackage{opex3,epsfig,subfigure,changebar,picinpar,rotating,graphicx,pstricks,ulem,epstopdf,amsmath,cite}

\begin{document}
\title{Quantum-correlated two-photon transitions to excitons in semiconductor quantum wells}
\author{L. J. Salazar,$^{1}$ D. A. Guzm\'an,$^{1,2}$ F. J. Rodr\'{\i}guez,$^{1}$ and L. Quiroga$^{1,*}$}
\address{$^{1}$Physics Department and $^{2}$Quantum Optics
Laboratory, Universidad de los Andes, A.A. 4976, Bogot\'a D.C.,
Colombia}
\email{$^{*}$lquiroga@uniandes.edu.co}
\begin{abstract}
The dependence of the excitonic two-photon absorption on the quantum correlations (entanglement) of exciting biphotons by a semiconductor quantum well is studied. We show that entangled photon absorption can display very unusual features
depending on space-time-polarization biphoton parameters and absorber density
of states for both bound exciton states as well as for unbound electron-hole pairs.
We report on the connection between biphoton entanglement,
as quantified by the Schmidt number, and absorption
by a semiconductor quantum well. Comparison between frequency-anti-correlated, unentangled and frequency-correlated
biphoton absorption is addressed. We found that exciton oscillator strengths are highly increased when photons arrive almost simultaneously in an entangled state. Two-photon-absorption becomes a highly sensitive probe of photon quantum correlations when narrow semiconductor quantum wells are used as two-photon absorbers.
\end{abstract}


\section{Introduction}
Quantum correlations (entanglement) with no classic analogs are among the most surprising consequences of quantum mechanics.
A multipartite quantum system in an entangled state must be described as a non-separable entity without any
reference to well defined individual states. Although, the measurement of a local observable can yield to completely random values,
there are strong correlations between different parts of the composite system.
These non-classical correlations can usually appear in the form of energy, linear momentum or angular momentum
conservation requirements. They are crucial for
quantum-enhanced technologies \cite{nielsen} and would give access to the implementation of new quantum algorithms. 
Widely studied entangled systems are those formed by quantum-correlated photons which are produced
by the so-called spontaneous-parametric-down-conversion (SPDC) mechanism \cite{shi} where a continuous or pulsed pump laser beam is sent to an optical nonlinear crystal. In a SPDC elemental process one laser pump photon is destroyed and two lower energy photons, signal and
idler, are created within the material. A signal-idler pair (or biphoton)
is a highly entangled system in variables such as energy, linear momentum, angular
momentum and polarization \cite{edamatsu}. In the last two decades,
this process has been exploited in different quantum optics
experimental scenarios especially designed for probing the
foundations of quantum physics \cite{mandel,zeilinger}.
Recently, the interfacing of such quantum optical beams with solid-state quantum objets, like Cooper-pairs in a superconductor box \cite{tian} and many-body correlations in semiconductors \cite{kira}, has proven to be an emerging fertile field worthwhile of investigation. 
In the present work the theoretical study of the absorption of quantum-correlated photons by electron-hole pairs in a semiconductor quantum well (QW) is addressed.
Of special interest is the understanding of the interplay between
field-correlation features and photon absorption. A well known
result is that the two-photon-absorption (TPA) process provides information about higher-order correlation
functions of the radiation field \cite{mollow,agarwal}. In particular, the photon first-order correlation function, although important in determining the spectrum of one-photon absorption response by a matter system, is of limited usefulness
for describing multi-photon processes. Therefore, in order to properly describe the matter optical response to new light sources, including quantum-correlated photons, the inclusion of higher-order correlation functions is crucial to theoretically get the correct optical response. This is a first point of interest in the present work. On the other hand, from an experimental point of view, several procedures have been established
to control the different entanglement features of signal/idler pairs. For instance, by inserting spectral filters in the paths of signal and idler
photons allows for the production of narrow-band time-entangled
biphotons. Additional control mechanisms affecting the
biphoton wavefunction shape, including the engineering of the
spatial profile of the pump laser and  selecting the geometry of the
nonlinear crystal\cite{dayan1,carrasco,valencia}, provide useful tools to manipulate the two-photon wavefunction or equivalently the photon temporal correlations. As a second goal of the present work, we perform a comparative study of the absorption response of
a semiconductor QW for differently quantum-correlated signal/idler photon pairs.

Previous entangled-two-photon-absorption (ETPA) by specific matter systems include simple molecules \cite{kojima}, dendrimer molecular systems \cite{goodson,goodson2}, organic
cromophores \cite{goodson3} and simple solid systems
\cite{lissandrin}. Furthermore, on the theory side, surprising new phenomena such as entanglement induced transparency
\cite{fei1,fei2,perina} and ETPA by two spatially distant atoms\cite{scully} have arisen great interest. However, all of these systems are intrinsically narrow-bandwidth absorbers.
In contrast, recent experimental investigations using bulk $GaAs$ show that semiconductors act as efficient large-bandwidth, TPA\cite{fabre1} and ETPA\cite{fabre2}, receptors. Semiconductor materials provide advantages for photon absorption over atomic systems due to the possibility of tailoring the band density of states and/or the density of carriers through the control of the confinement dimensionality and excitation mechanisms. Thus, they offer the opportunity of measuring photon temporal correlations in a femtosecond timescale by exploiting  processes where the almost simultaneous arriving of two photons is required. It is therefore natural to ask whether more versatile large-bandwidth devices could be
engineered by resorting to semiconductor nanostructures like a QW as the two-photon receptor. Large optical nonlinearities have been measured in
such systems, with a wealth of new optical phenomena which are difficult to observe in bulk crystals.
In particular, photon absorption in such nanostructures
turns out to be a nonmonotonic function of the photon energy with
steps occurring whenever transitions are available between new pairs
of quantized subbands. In the quantum limit of strong confinement
($L\sim$ few nanometers) those spectra present a very rich
structure showing new features associated to a transformed density
of available states as well as to inter-particle interaction
effects. Furthermore, two-photon emission from semiconductor QWs has been realized to be a source of high-intensity room-temperature entangled photons\cite{hayat1}. Since ETPA
processes provide information about higher order correlation functions of the
radiation field, QWs nanostructures may be therefore useful for in-situ
detection of entangled photons produced by other on-chip
semiconductor sources like QWs\cite{hayat1} or quantum dots\cite{shields1}.

We limit ourselves to the case when one-photon absorption is
excluded. This paper is organized as follows: In Section II we
briefly review the main aspects of quantum-correlated photons and the general ETPA formalism for both non-interacting electron-hole pairs and excitons in QWs. In Section III results of ETPA by QWs are discussed. Finally, in Section IV the main conclusions of the present work are summarized.

\section{Theoretical background}
\begin{figure}[tbh]
\begin{center}
\includegraphics[height=5.5cm,width=9.5 cm] {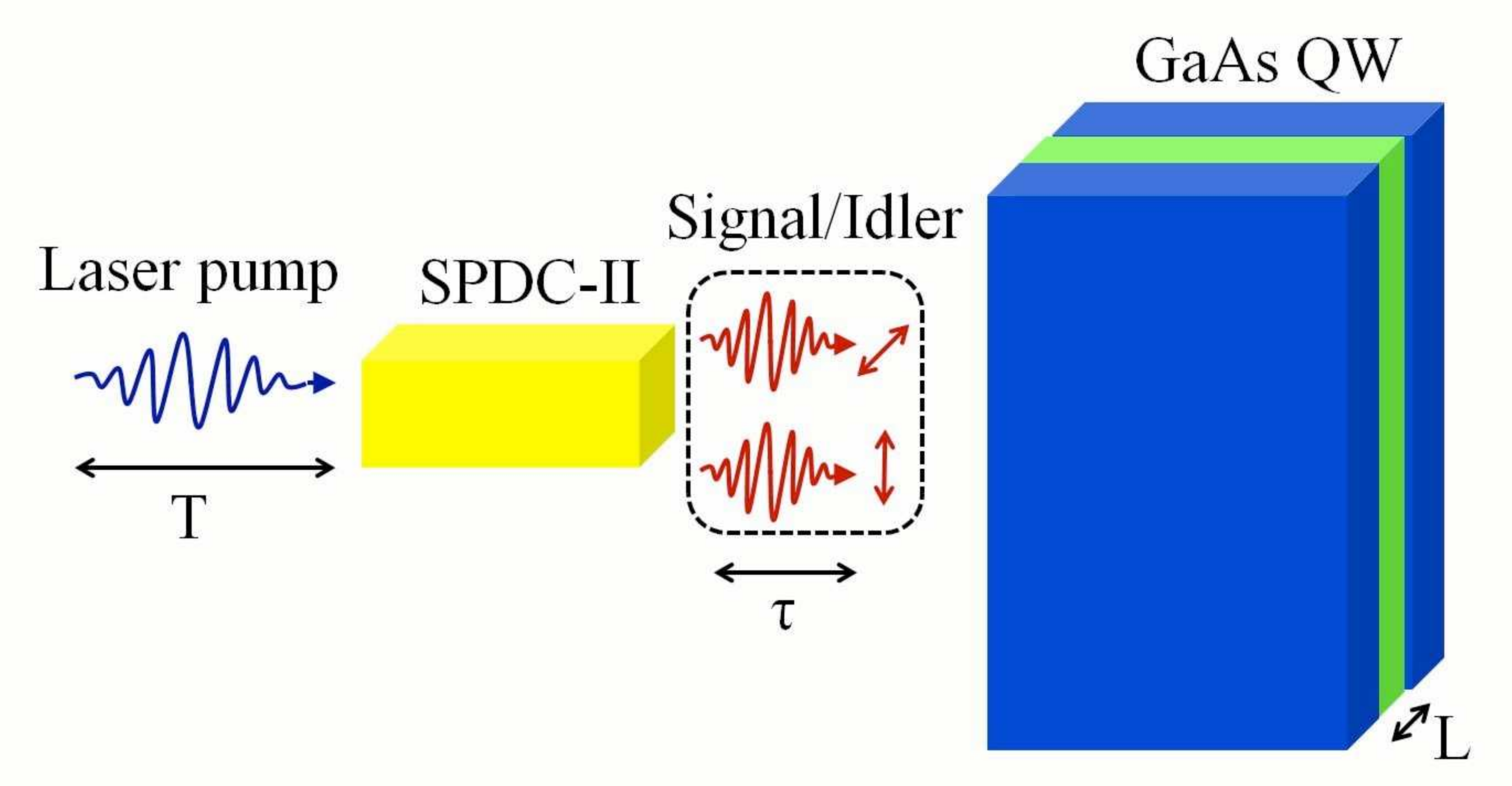}
\end{center}
\caption{SPDC-II produced biphotons shining a semiconductor QW. Orthogonal signal/idler polarizations are shown. $T$ denotes the laser pump duration whereas $\tau$ is associated with the entanglement-time of the signal/idler pair as generated by the SPDC nonlinear crystal.} \label{fig0}
\end{figure}

The ETPA probability by a semiconductor QW requires a full quantum description of both the radiation field and the matter system. The radiation-matter system of interest is outlined in Fig. \ref{fig0}. We first address the issue concerning the generation and description of quantum correlated biphotons. Secondly, the description of the absorber system, a semiconductor QW, is provided by focusing on the selection rules governing the interaction of radiation-matter of interest for the present work. Explicit expressions are finally given for the ETPA rate in both cases of no excitonic effects as well as when excitons are included.

\subsection{Quantum-correlated biphoton wavefunctions}
We start by describing the quantum-correlated photon wavefunctions in both time and frequency variables. For the purposes of concreteness we shall focus on realistic entangled photon sources based on type-II SPDC processes\cite{jptorres} for which there exists the possibility of experimentally tailoring the degree of quantum correlation among the photons, allowing even to produce uncorrelated twin photons.
In this way, it becomes possible to analyze, within the same theoretical framework, the novel aspects that photon quantum correlations bring to the basic process of absorption by a semiconductor nanostructure.The biphoton wavefunction is given by
\begin{eqnarray}
\psi(t_1,t_2)=\langle 0|\hat{a}_2(t_2)\hat{a}_1(t_1)|\Psi_2 \rangle
\label{Eq:p2}
\end{eqnarray}
where $|\Psi_2 \rangle$ is the radiation quantum state emerging from the type-II SPDC device while  $\hat{a}_1$ and $\hat{a}_2$ denote annihilation operators for signal/idler photons with corresponding field amplitudes ${\cal E}_1$ and ${\cal E}_2$, respectively (${\cal E}_i$ has dimensions of electric field times square root of time). 
\begin{figure}[tbh]
\vspace{-1cm}
\begin{center}
\includegraphics[scale=1.9,height=6cm, width=5.0cm, bb=0 0 380 350] {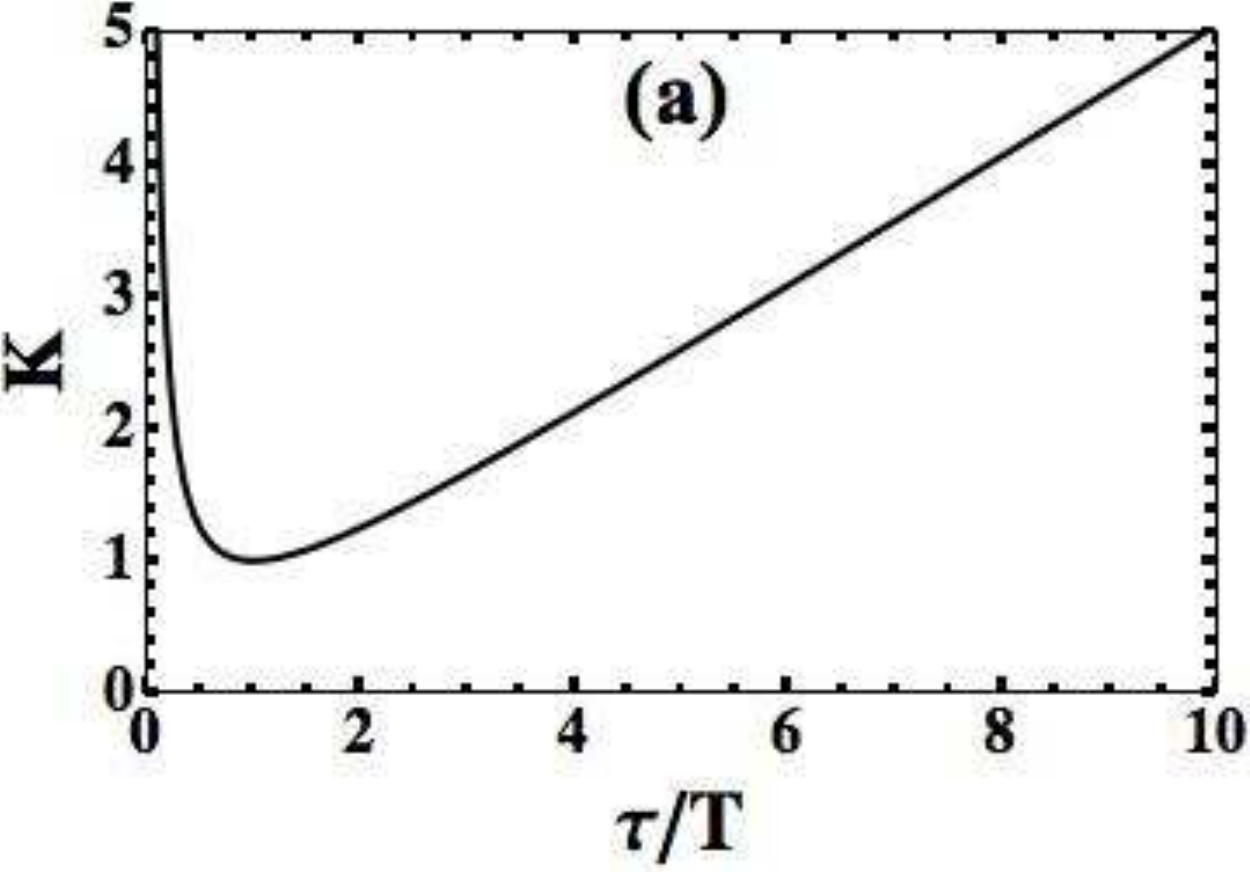}
\includegraphics[scale=1.9,width=5.0 cm,height=4.4 cm,bb=0 0 380 350] {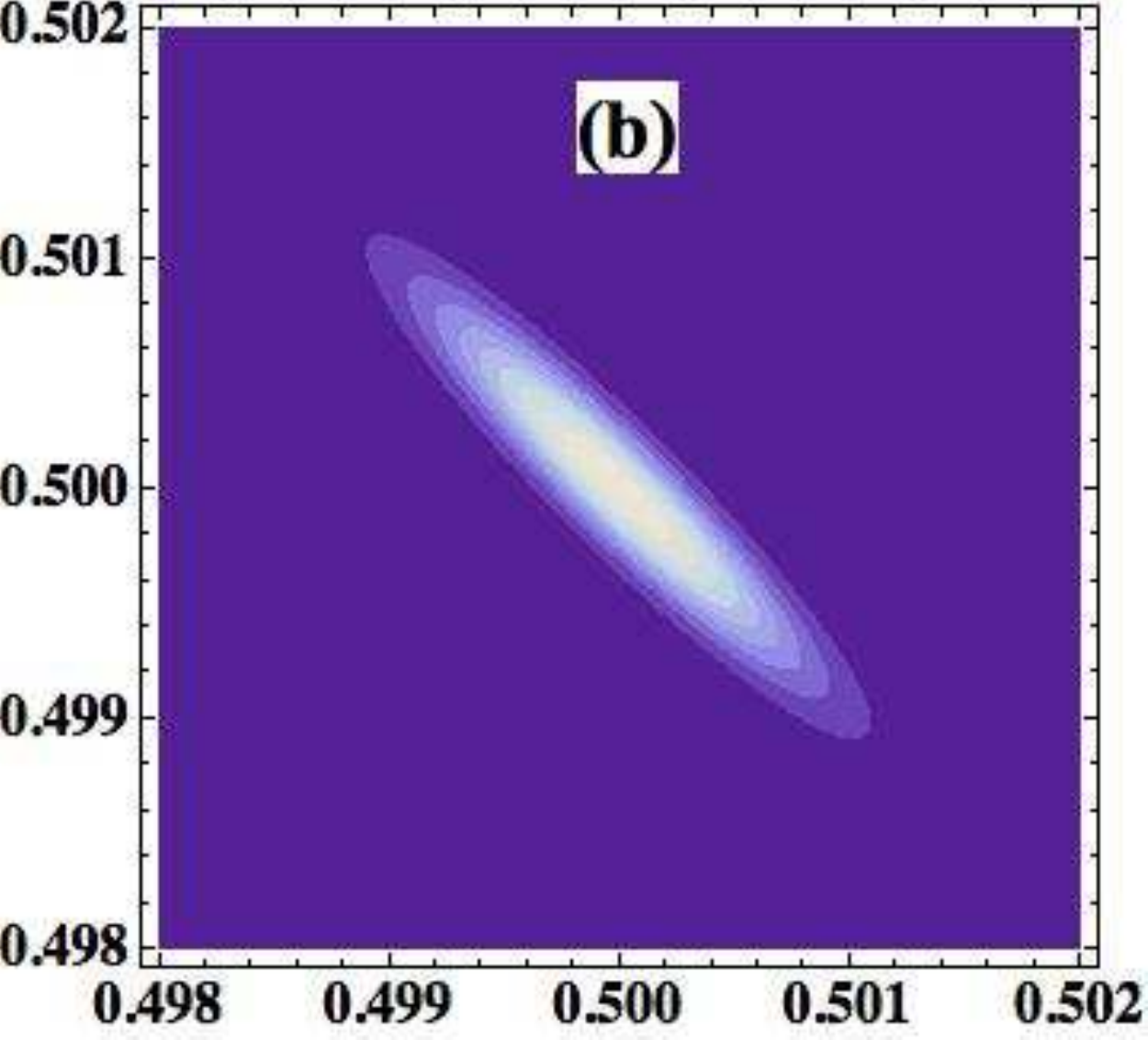}
\includegraphics[scale=1.9,width=5.0 cm,height=4.5 cm,bb=0 0 380 350] {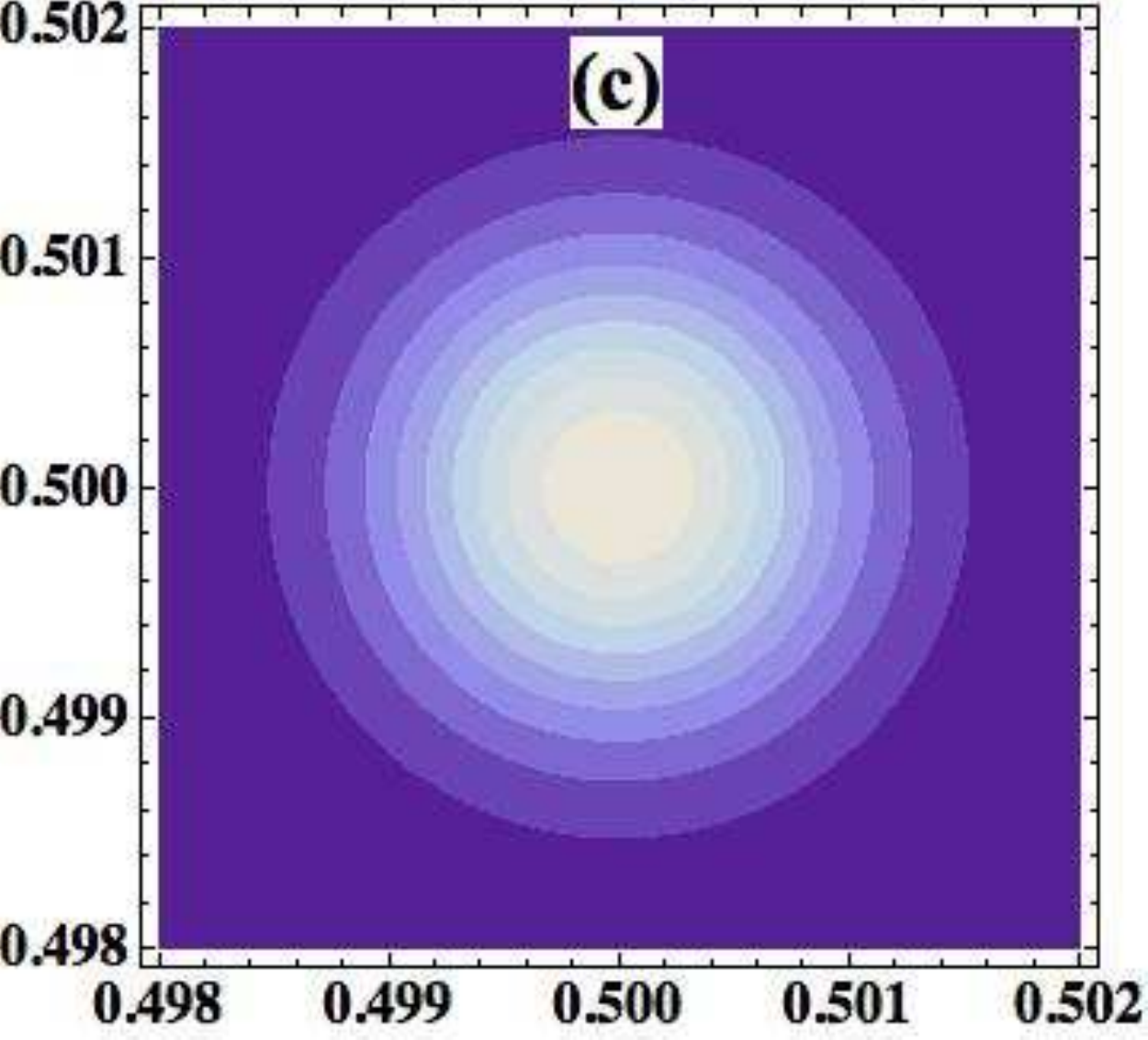}
\includegraphics[scale=1.9,width=5.0 cm,height=4.4 cm,bb=0 0 380 350] {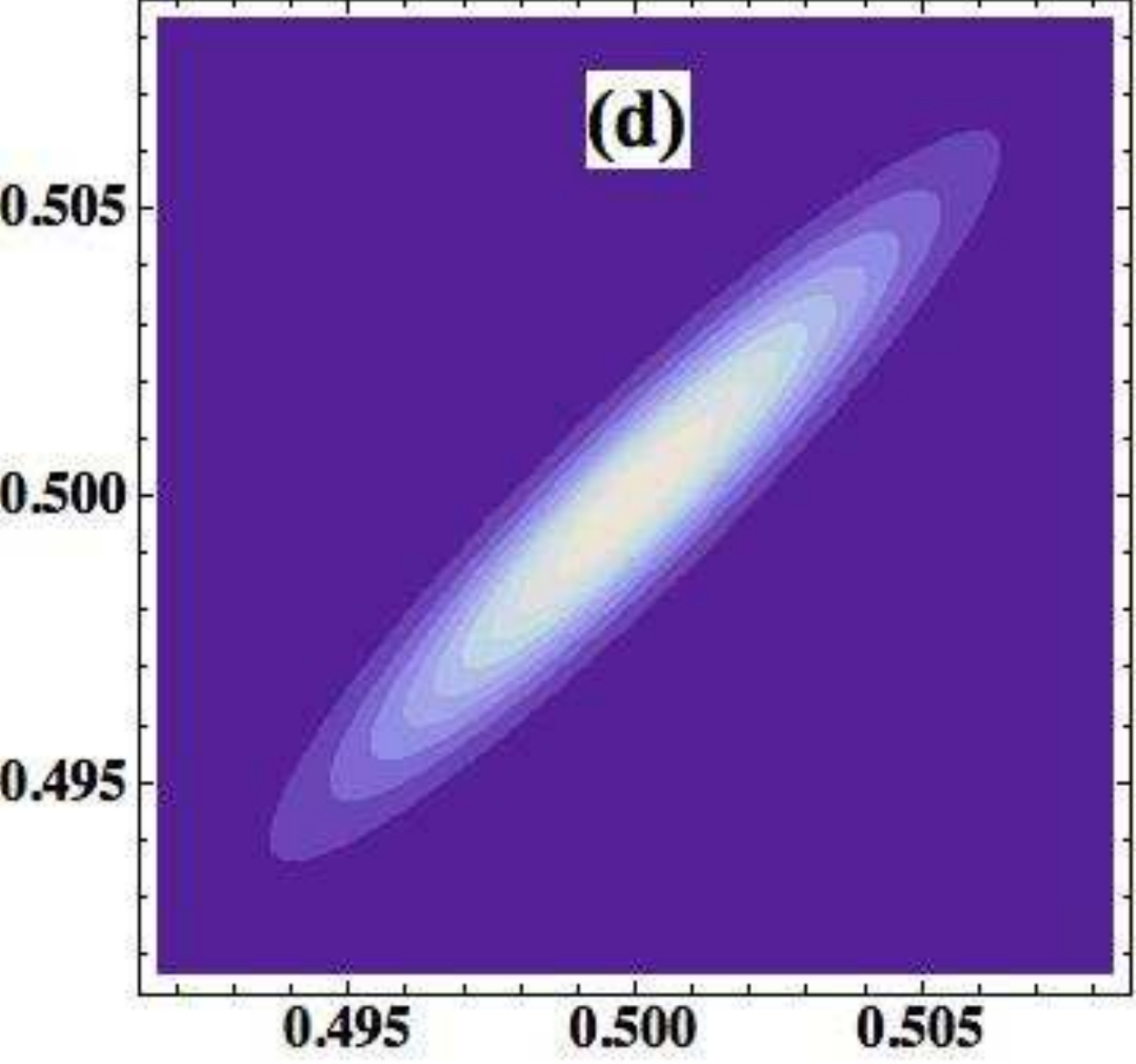}
\end{center}
\caption{(a) Biphoton entanglement measure, Schmidt $K$ number, as a function of $\frac{\tau}{T}$. (b), (c) and (d): Joint spectral intensity from Eq. (\ref{Eq:p13}). The Schmidt number for graphs (b) and (d) is the same, $K=3$. In (b) $\frac{\tau}{T} < 1$ corresponding to a a FAC biphoton while in (d) $\frac{\tau}{T} > 1$ related to a FC biphoton. For
graph (c) the Schmidt number is $K=1$ corresponding to an unentangled biphoton state. Axes display the signal and idler frequencies in units of $\omega_p^{(0)}$ with $\omega_p^{(0)} \tau=500$.} \label{fig1}
\end{figure}

The Fourier transform of the type-II entangled biphoton wavefunction\cite{shi} has the form
\begin{eqnarray}
\Psi(\omega_1,\omega_2)&=&{\bar{N}}e^{-(\omega_1+\omega_2-\omega_p^{(0)})^2T^2}
\Phi\left ( \omega_1-\omega_2-\delta\omega \right )
\label{Eq:p12}
\end{eqnarray}
where ${\bar{N}}$ is a normalization constant which ensures that $\int d\omega_1\int d\omega_2|\Psi(\omega_1,\omega_2)|^2=1$, $\omega_p^{(0)}=\omega_s+\omega_i$ is the central pump frequency, $\omega_s$ ($\omega_i$) the central signal (idler) frequency, $T$ is the pump coherence time (inverse of pump bandwidth) and $\delta\omega=\omega_s-\omega_i$.
In a great variety of cases, especially when the relation $2v_p^{-1}=v_s^{-1}+v_i^{-1}$ holds ($v_j$ with $j=p,s,i$ denoting group velocities of pump, signal and idler photons, respectively), the Fourier transformed biphoton wavefunction, Eq. (\ref{Eq:p12}), can be assumed to have a double-gaussian bipartite form\cite{jptorres,fedorov1, fedorov2}
\begin{eqnarray}
\Psi(\omega_1,\omega_2)={\bar{N}}e^{-(\omega_1+\omega_2-\omega_p^{(0)})^2T^2}e^{-(\omega_1-\omega_2-\delta\omega)^2\tau^2}
\label{Eq:p13}
\end{eqnarray}
with ${\bar{N}}=2\sqrt{\tau T/\pi}$ and the effective signal/idler quantum-correlation time is represented by $\tau$. While different values of $\tau$ can be attained by modifying non-linear crystal properties such as its length, and group velocities of different pump/signal/idler photons, the values of $T$ are essentially associated to coherence properties of the exciting laser pump, thus being affected by the pump laser pulse duration, among other factors.

Thus, quantum-correlated photons produced by SPDC devices can be designed with a high spectral purity for sufficiently large values of $T$, and at the same time with a narrow time-coincidence window as characterized by small values of $\tau$. This last feature marks a strong departure from the behavior of classical (laser) light.

The biphoton wavefunction given in Eq. (\ref{Eq:p13}) encloses different photon quantum correlations which can be controlled by changing the pump bandwidth (equivalently the pump pulse duration $T$) and the correlation time $\tau$. These two experimentally accessible control parameters lead directly to a measure of the amount of entanglement of two photon modes as quantified by the so-called Schmidt number, which turns out to be\cite{fedorov2}
\begin{eqnarray}
K=\frac{1}{2}\left ( \frac{\tau}{T}+\frac{T}{\tau} \right )
\label{Eq:p15}
\end{eqnarray}
It reaches the minimum value of $1$ (unentangled biphoton state) when $T=\tau$ as shown in Fig. \ref{fig1}(a). Although, identical values of the Schmidt number can be obtained for two different ratios $\tau/T$, the quantum correlation features can differ significantly. This point is illustrated in Fig. \ref{fig1}(b)-\ref{fig1}(d) where the biphoton joint spectral intensity is plotted for different pump pulse duration. Figs. \ref{fig1}(b) and \ref{fig1}(d) correspond to identical biphoton entanglement, $K=3$. However, while in Fig. \ref{fig1}(b) a frequency-anti-correlated (FAC) biphoton is displayed for long pump pulses, $\tau/T<1$, in the opposite case, short pump pulses, $\tau/T>1$, the biphoton state as shown in Fig. \ref{fig1}(d) is frequency-correlated (FC). The uncorrelated biphoton joint intensity, $K=1$, produced by a pump pulse with $T=\tau$, is shown in Fig. \ref{fig1}(c).

\subsection{ETPA: no-excitonic effects}
Now, we describe how quantum-correlated photons interact with a semiconductor nanostructure. We restrict ourselves to consider a semiconductor infinite square well potential of thickness $L$ in direction $z$. The initial state corresponds to the QW in the ground state $|G \rangle$ and the radiation field with two quantum-correlated photons described by the state $|\Psi_2\rangle$ as produced by a SPDC device, while the final state corresponds to the QW in any excited state $|n_c,n_v,\vec{k}\rangle$ with an electron in conduction sub-band $n_c$, a hole in the valence sub-band $n_v$ (identical two-dimensional $\vec{k}$ wavevectors for electron and hole, as discussed below) and the photon field in the vacuum state, $|0\rangle$.

\begin{figure}
\hspace{3cm}
\hspace{0cm}
\includegraphics[height=8.0cm,width=8.5 cm] {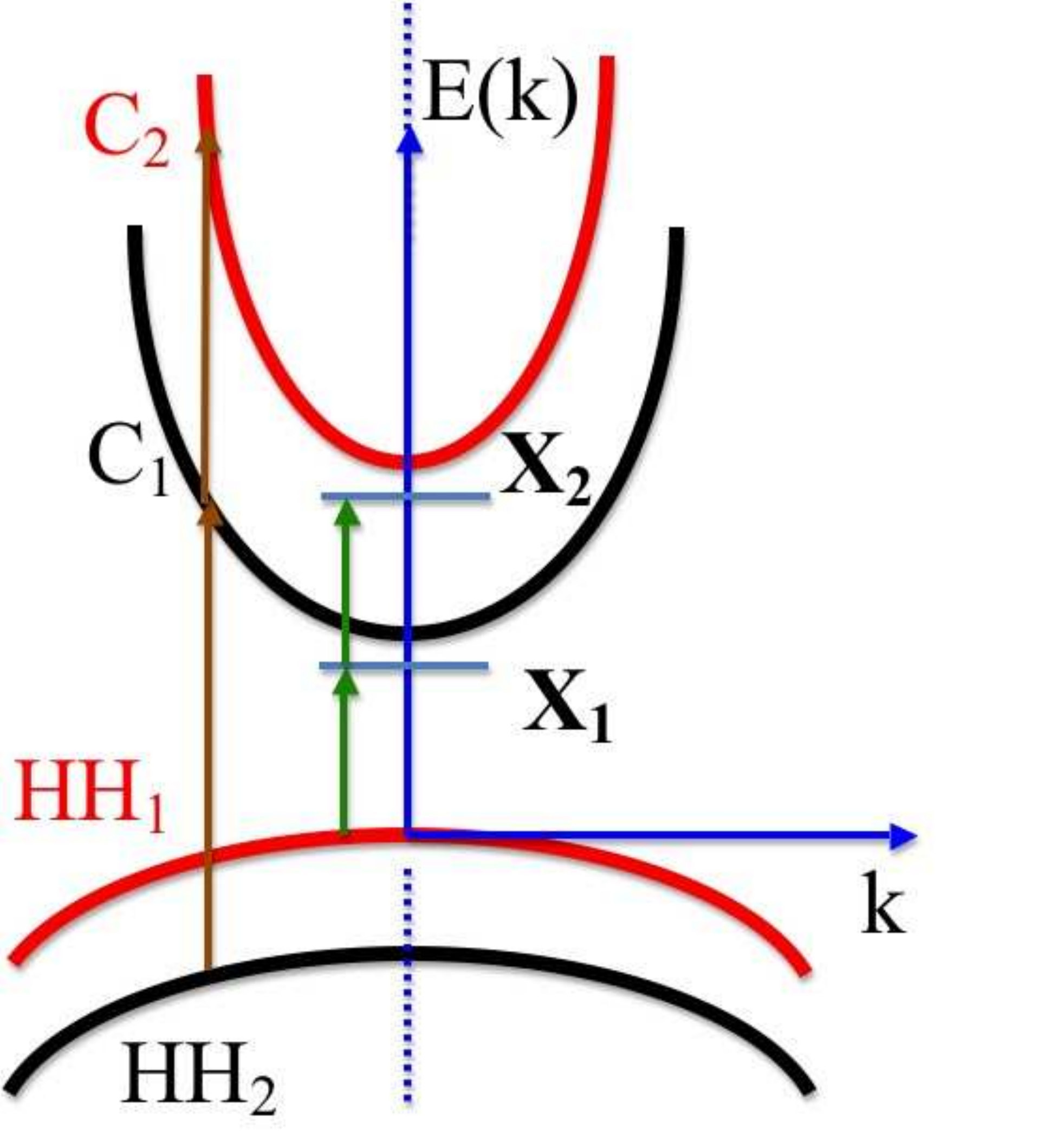}
\vspace{0cm}
\caption{Schematic diagram of energy levels  for a semiconductor quantum well. Two photon  transitions involving  electron-hole states (brown arrows) and
exciton states (green arrows) are depicted. $C_{i}$ is the $i-th$ conduction sub-band, $HH_{i}$ is the $i-th$ heavy hole valence sub-band and $X_i$ denote discrete exciton states. } 
\label{bandstructure1}
\end{figure}

The ETPA amplitude\cite{kitano} is given by $\alpha_2(n_c,n_v,\vec{k})=\langle n_c,n_v,\vec{k},0|\hat{U}_2|G,\Psi_2\rangle$ where the evolution operator is evaluated up to second order in the radiation-matter coupling Hamiltonian $\hat{H}_I(t)$ (in the interaction picture), corresponding to $\hat{U}_2=-\frac{1}{\hbar^2}\int_{-\infty}^{\infty} dt_2\int_{-\infty}^{\infty} dt_1\hat{H}_I(t_2)\hat{H}_I(t_1)\theta(t_2-t_1)$, where $\theta(t)$ is the Heaviside step function
 \begin{eqnarray*} 
\theta (t) =
  \begin{cases}
  1 & t \geq 0\\
  0 &  t < 0\\
  \end{cases} 
\end{eqnarray*} 
The ETPA probability amplitude turns out to be
\begin{eqnarray}
 \alpha_2(n_c,n_v,\vec{k})=-\frac{i8}{\pi^2}\frac{P_{cv}L}{\hbar m_0 E_g}e^2{\cal E}_1{\cal E}_2\frac{n_cn_v}{(n_c^2-n_v^2)^2}
\int_{-\infty}^{\infty}dt_2\int_{-\infty}^{\infty} dt_1\theta(t_2-t_1)e^{i \Omega_2t_2}e^{i \Omega_1t_1}\psi(t_1,t_2)
\label{Eq:p1}
\end{eqnarray}
where the biphoton wavefunction in time domain, $\psi(t_1,t_2)$, is given by Eq. (\ref{Eq:p2}), the interband transition energy corresponds to
\begin{eqnarray}
\hbar\Omega_1=Eg+\frac{\hbar^2 \pi^2}{2\mu L^2}n^2+\frac{\hbar^2 k^2}{2\mu}\, ,\, n=min\{n_c,n_v \}
\label{Eq:p91}
\end{eqnarray}
and the intraband transition energy can be one of the following
\begin{eqnarray}
\nonumber \hbar\Omega_2&=&\frac{\hbar^2 \pi^2}{2m_c^*L^2}(n_c^2-n_v^2)\, ,\, n_c > n_v\\
\hbar\Omega_2&=&\frac{\hbar^2 \pi^2}{2m_v^*L^2}(n_v^2-n_c^2)\, ,\, n_v > n_c
\label{Eq:p92}
\end{eqnarray}
In the last two equations, the first one describes an intraband conduction transition whereas the second one is associated to an intraband valence transition.
In Eqs. (\ref{Eq:p1})-(\ref{Eq:p92}) $m_0$ is the free electron mass, $m_c^*$ ($m_v^*$) the effective electron (hole) mass, $\mu$ denotes the reduced electron-hole mass and $E_g$ is the bulk semiconductor energy gap. The $\vec{r}$-gauge for the dipole transition elements have been adopted, which read as $\langle n_c,\vec{k},n_v,\vec{k}^{\prime}|\hat{u}_{\epsilon}\cdot e\hat{\vec{r}}|G\rangle=
-\frac{ie\hbar}{m_0E_g}(\hat{u}_{\epsilon}\cdot \vec{P}_{cv})\delta_{n_c,n_v}\delta_{\vec{k},\vec{k}^{\prime}}$
for direct interband transitions with $\vec{P}_{cv}$ the interband Kane transition element, and $\langle n_{\nu},\vec{k}|\hat{u}_{z}\cdot e\hat{\vec{r}}|\tilde{n}_{\nu},\vec{k}^{\prime}\rangle=
-\frac{8e L}{\pi^2}\frac{n_{\nu}\tilde{n}_{\nu}}{(\tilde{n}_{\nu}^2-n_{\nu}^2)^2}\delta_{\vec{k},\vec{k}^{\prime}}$
for intraband transitions within the $\nu=c$ conduction band or the $\nu=v$ valence band. For intraband transitions a selection rule, imposing a change of parity, implies the condition $|\tilde{n}_{\nu}-n_{\nu}|=1,3,5...$ should hold\cite{spector1,pasquarello1}. In these expressions $\hat{u}_{\epsilon}$ denotes an unitary vector pointing in directions $x$, $y$ or $z$
and the allowed transistions are schematically depicted in Fig. \ref{bandstructure1}.
Quantum correlated photons emitted by a type-II SPDC crystal have orthogonal polarizations (see Fig. \ref{fig0}). Consequently, we assume the inter-band transition is induced by an on-plane polarized photon, i.e. $\hat{u}_{\epsilon}=\hat{u}_{x}$, while intra-band transitions respond to perpendicularly polarized photons, i.e. $\hat{u}_{\epsilon}=\hat{u}_{z}$. In this way, TPA from a heavy-hole in the initial ground state is possible when the interband transition occurs first excited by the $\hat{u}_{x}$ polarized photon followed by the intra-band transition induced by the $\hat{u}_{z}$ polarized photon. Thus, a heavy-hole TPA, which is a forbidden transition\cite{pasquarello1} under laser excitation with identical $\hat{u}_{z}$ photon polarizations, is now permitted for type-II SPDC generated biphotons. Therefore, inter-band transitions involving heavy holes are the only ones we shall consider below.

Given the time ordering on the photon induced transitions, it turns out to be more appropriate\cite{kitano} to define a modified Fourier transformed biphoton wavefunction as (note that $\tilde{\Psi}(\omega_1,\omega_2)$ has dimensions of time) $\tilde{\Psi}(\omega_1,\omega_2)=\frac{1}{2\pi}\int_{\infty}^{\infty}dt_2\int_{\infty}^{\infty} dt_1\theta(t_2-t_1)\psi(t_1,t_2)
e^{i\omega_2t_2}e^{i\omega_1t_1}$. For the frequency biphoton wavefunction given by Eq. (\ref{Eq:p13}) we obtain
\begin{eqnarray}
\tilde{\Psi}(\omega_1,\omega_2)=\sqrt{\frac{\tau T}{\pi}}e^{-(\omega_1+\omega_2-\omega_p^{(0)})^2T^2}F\left [ (\omega_1-\omega_2-\delta\omega)\frac{\tau}{2} \right ]
\label{Eq:p9}
\end{eqnarray}
with $F(z)=e^{-z^2}+\frac{2i}{\sqrt{\pi}}D(z)$ and $D(z)=\int_0^ze^{-(z^2-y^2)}dy$ the Dawson function. Thus, the ETPA amplitude, Eq. (\ref{Eq:p1}), reads as
\begin{eqnarray}
\alpha_2(n_c,n_v,\vec{k})=-\frac{i16}{\pi}\frac{P_{cv}L}{\hbar m_0 E_g}e^2{\cal E}_1{\cal E}_2\frac{n_cn_v}{(n_c^2-n_v^2)^2}
\tilde{\Psi}(\Omega_1,\Omega_2)
\label{Eq:p6}
\end{eqnarray}
with $\Omega_1$ ($\Omega_2$) given by Eq. (\ref{Eq:p91}) (Eq. (\ref{Eq:p92})) and $\tilde{\Psi}(\Omega_1,\Omega_2)$ as written in Eq. (\ref{Eq:p9}). The condition $|n_c-n_v|=1,3,5...$ should be fulfilled. It is evident from Eq. (\ref{Eq:p6}) that ETPA efficiency becomes highly dependent on the mapping of the absorber transition energies, $\Omega_1$ and $\Omega_2$, onto the biphoton signal/idler frequencies.

Finally, the ETPA probability is the sum over final states as ${\cal P}_2(\omega)=\sum_{n_c}\sum_{n_v} \sum_{\vec{k}}|\alpha_2(n_c,n_v,\vec{k})|^2$, which for the present case turns out to be
\begin{eqnarray}
{\cal P}_2(\omega)=\Delta\sum_{n=1}^{\infty}\sum_{\tilde{n}=1,3,5...}\sum_{\nu=c,v}\left [ \frac{n(n+\tilde{n})}{(\tilde{n}(2n+\tilde{n}))^2}\right ]^2
\nonumber
\\
\int_0^{\infty}d\kappa\, \kappa
\left | \tilde{\Psi}\left (1+\alpha (n^2+\kappa^2),\alpha \tilde{n}(2n+\tilde{n})\frac{\mu}{m_{\nu}^*} \right ) \right |^2
\label{Eq:p7}
\end{eqnarray}
Dimensionless variables have been used: energies in units of the bulk semiconductor gap $E_g$, frequencies in units of $\omega_g=E_g/\hbar$, time in units of $\omega_g^{-1}$, $\alpha=\frac{1}{E_g}\frac{\hbar^2 \pi^2}{2\mu L^2}$, $\kappa=k\frac{L}{\pi}$ and $\Delta=\left [ \frac{16 P_{cv}}{m_0E_g^2}e^2{\cal E}_1{\cal E}_2 \right ]^2\frac{A_{QW}}{2\pi}$, being $A_{QW}$ the QW transversal area.
In case the biphoton wavefunction is unentangled (see Fig. \ref{fig1}(c)) the classical laser absorption results\cite{pasquarello1,spector1} are easily recovered.

\subsection{ETPA: excitonic effects}
The above results, can be easily extended to the case where excitonic effects are taken into account. The interband transition occurs now between the semiconductor ground state $|G\rangle$ and an exciton state denoted by $|n_c,n_v,\beta,m\rangle$ where $n_c$ and $n_v$ represent as before sub-band indices for electron and hole, respectively, while the relative exciton bound state is marked by the principal quantum number, $\beta$, and the angular momentum projection number in direction $z$ is $m$. The matrix element for this transition becomes now\cite{pasquarello2} $\langle n_c,n_v,\beta,m|\hat{u}_{\epsilon}\cdot e\hat{\vec{r}}|G\rangle=
-\frac{ie\hbar}{m_0E_g}(\hat{u}_{\epsilon}\cdot \vec{P}_{cv})\sqrt{A_{QW}}\Phi_{\beta,m}(\vec{r}=0)\delta_{n_c,n_v}\delta_{m,0}$ where $a_B$ denotes the 3D Bohr radius and $\Phi_{\beta,m}(\vec{r})$ is the exciton wavefunction. As indicated by this selection rule, only interband transitions producing $m=0$ or S-excitons are allowed. On the other hand, the intra-band matrix element becomes now\cite{pasquarello2} $\langle n_c^{\prime},n_v^{\prime},\beta^{\prime},m^{\prime}|\hat{u}_{\epsilon}\cdot e\hat{\vec{r}}|n_c,n_v,\beta,m\rangle=-\frac{8e L}{\pi^2}\left (\frac{a_B}{L}\right )^2\frac{n_{\nu}n_{\nu}^{\prime}}{(n_{\nu}^{\prime 2}-n_{\nu}^2)^2}\delta_{\beta,\beta^{\prime}}\delta_{m,m^{\prime}}$
for intraband transitions within the $\nu=c$ conduction band or the $\nu=v$ valence band.

The excitonic ETPA probability becomes
\begin{eqnarray}
\nonumber {\cal P}_2(\omega)&=&\Delta\sum_{n=1}^{\infty}\sum_{\tilde{n}=1,3,5...}\sum_{\nu=c,v}\left [ \frac{n(n+\tilde{n})}{(\tilde{n}(2n+\tilde{n}))^2}\right ]^2
\nonumber
\\
& &\left [ \frac{a_B^4}{L^2}\left | \Phi_{1S}(\vec{r}=0)\right |^2
\left | \tilde{\Psi}\left ( 1+\alpha n^2-\alpha_{1S},\alpha \tilde{n}(2n+\tilde{n})\frac{\mu}{m_{\nu}^*} \right )\right |^2\right. 
\nonumber
\\
&+&\left. \int_0^{\infty}d\kappa\, \kappa\left | \Phi_{e-h}(\vec{\kappa},\vec{r}=0)\right |^2
 \left | \tilde{\Psi}\left (1+\alpha (n^2+\kappa^2),\alpha \tilde{n}(2n+\tilde{n})\frac{\mu}{m_{\nu}^*} \right ) \right |^2\right ]
\label{Eq:p8}
\end{eqnarray}
where $\Phi_{1S}(\vec{r})$ and $\Phi_{e-h}(\vec{\kappa},\vec{r})$ represent the bound $1S$ exciton wavefunction and the unbound, but Coulomb interacting, electron-hole pair wavefunction, respectively. We limit ourselves to consider only the exciton ground-state, or $1S$-state, with dimensionless binding exciton energy denoted as $\alpha_{1S}=E_{1S}/E_g$.

Excitonic effects in a finite thickness QW can be described by relying upon a variational treatment\cite{bastard1} with a trial bound exciton wavefunction for the transverse relative motion such as $\Phi_{1S}(\vec{r})=\sqrt{\frac{2}{\pi}}\frac{1}{\lambda}e^{-\frac{r}{\lambda}}$, a hydrogen-like wavefunction with $\lambda$ the variational parameter,
which reproduces well the 2D limit corresponding to an infinitely narrow QW exciton. In this limit, $\lambda=a_B^{(2D)}=\frac{2\pi\epsilon \hbar^2}{\mu e^2}$ the $2D$ exciton Bohr radius\cite{chemla} ($a_B^{(2D)}=a_B/2$) and $E_{1S}^{(2D)}=\frac{\hbar^2}{2\mu a_B^{(2D)2}}>0$ is the $1S$-exciton binding energy ($\epsilon$ is the bulk semiconductor dielectric constant). For $L>0$ QWs, $\lambda$ should be obtained numerically by minimizing the exciton energy.
For unbound states, the
electron-hole wavefunction evaluated at the origin, second line of Eq. (\ref{Eq:p8}), produces the well known Sommerfeld enhancement factor\cite{chemla} given by $\left | \Phi_{e-h}(\vec{k},\vec{r}=0)\right |^2=S(k)=\frac{2}{1+e^{-\frac{2\pi}{k \lambda}}}$.
Note that $1 \leq S(k) \leq 2$ accounts for the effects of the electron-hole Coulomb interaction in the unbound states, taking its maximum value, $S(k)=2$ at the bottom edge of sub-bands, i.e. at $k=0$.
In the limit of no Coulomb interaction between electrons and holes, i.e. $\lambda\rightarrow \infty$, results following from Eq. (\ref{Eq:p7}) are obtained again.
In case the biphoton wavefunction is unentangled (see Fig. \ref{fig1}(c)) and $T\rightarrow \infty$, the classical laser absorption results\cite{shimizu1} are also recovered.

\section{Results}
In order to illustrate the results about ETPA by a QW we rely on recent experimental data where the quantum-correlated photons are produced using non-linear materials like periodically poled $LiNbO_3$ crystals\cite{fabre2} and $AlGaN$ Bragg reflection waveguides\cite{jptorres}, for which the central pump frequency is on the order $\hbar \omega_p^{(0)}\approx 1.5-1.6$ eV and the quantum correlation time, $\tau$, usually in the femtosecond timescale, varies typically in a range such as $50 \leq \omega_p^{(0)} \tau \leq 600$. We limit ourselves to the case of the semiconductor $GaAs$ as the absorber system. Thus, we take the following matter parameters\cite{harrison1}: $E_g=1.5$ eV, $m_c^*=0.07 m_0$, $m_v^*=0.5 m_0$, 3D exciton Bohr radius $a_B=116 {\AA}$, 3D exciton binding energy $E_{1S}^{(3D)}=4.8$ meV and different thicknesses $50 {\AA}\leq L \leq 200 {\AA}$. In all the results shown below, the ETPA signal shall be identified as ${\cal P}_2(\omega)/\Delta$, a dimensionless quantity. Furthermore, we concentrate in the signal/idler degenerate case $\omega_s=\omega_i=\omega_p^{(0)}/2=\omega$, and take $\omega$ as a free variable.

\begin{figure}[tbh]
\vspace{-3.5cm}
\hspace{-1cm}
\includegraphics[height=15.0cm,width=14.5 cm] {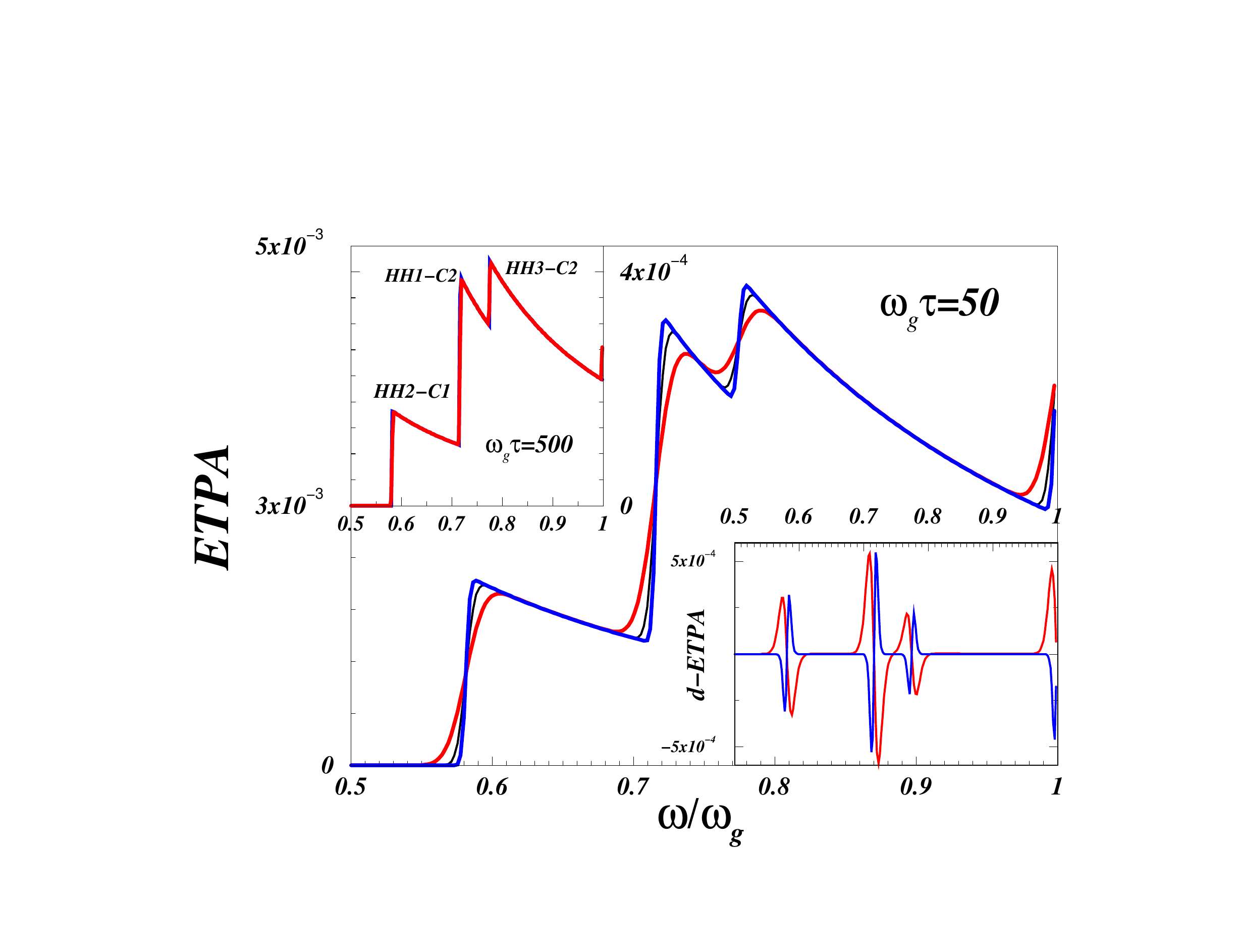}
\vspace{-2cm}
\caption{ ETPA, as a function of $\omega/\omega_g$, by a $L=a_B/2$ QW when excitonic effects are not included. Main graph corresponds to highly quantum correlated
biphotons with $\omega_g \tau=50$. Blue curve describes frequency-anti-correlated biphotons with $K=1.3$. Black curve denotes unentangled biphotons, $K=1$. Red curve corresponds to frequency-correlated biphotons with $K=1.3$. Upper inset: ETPA from weakly quantum correlated
biphotons with $\omega_g \tau=500$. Bottom inset: Differential ETPA for $K=1.3$ biphotons: FAC, blue curve; FC, red curve.} \label{fig2}
\end{figure}
First, we discuss ETPA results when Coulomb effects are ignored. As can be seen in Fig. \ref{fig2}, for a narrow QW with $L=a_B/2$, the main effects of photon quantum correlations are evident in variations of the absorption edges. Steps in this plot occur when the degenerate signal/idler biphoton is on resonance with the transitions (from low to high energy): HH2-C1, HH1-C2 and HH3-C2, respectively, where HH denotes a heavy-hole sub-band while C denotes a conduction sub-band. This type of TPA transitions is forbidden when both photons are $z$-polarized, but here these transitions are allowed by the fact that type-II SPDC produces cross-polarized photons, breaking that selection rule.  ETPA for two extreme values of the signal/idler correlation time $\tau$, with the same entanglement Schmidt measure $K=1.3$, are shown in Fig. \ref{fig2} (for comparison, it is also depicted the unentangled case, $K=1$). 
\begin{figure}[tbh]
\vspace{-1cm}
\hspace{-1cm}
\includegraphics[height=15.0cm,width=15.5 cm] {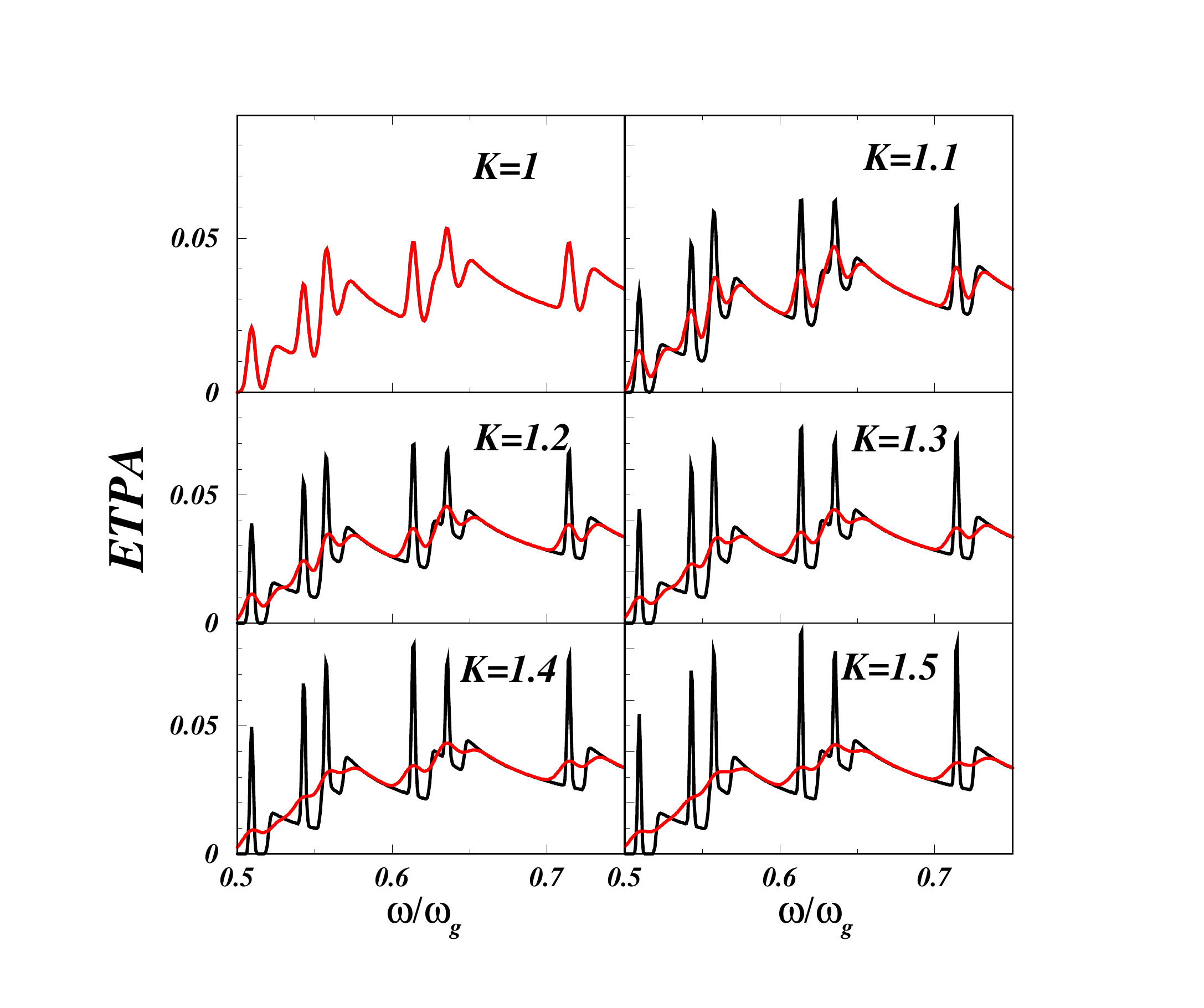}
\vspace{-1.5cm}
\caption{ ETPA by a $L=a_B$ QW with excitonic effects included, for highly quantum correlated biphotons, $\omega_g \tau=100$ and different photon entanglement, $K$. In each plot, FAC (black curves) and FC (red curves). For $K=1$ red and black curves are the same. } \label{fig3}
\end{figure}
For highly quantum correlated photons, main graph in Fig. \ref{fig2}, $\omega_g \tau=50$, changes in the absorption edge for each allowed two-photon transition are different depending on the frequency-anti-correlated or frequency-correlated biphoton feature. These differences are better observed in the bottom inset where differential-ETPA (d-ETPA, defined as $ETPA(K\neq 1)$-$ETPA(K=1)$) is shown. As compared with the unentangled signal, FAC biphotons yield to an entanglement-induced two-photon transparency\cite{fei1} (d-ETPA negative) in the low-energy side of the edge steps, while FC biphotons produce the reverse effect (d-ETPA positive), i.e. an entanglement-induced two-photon opacity in the same energy sector. These behaviors exchange roles on the high energy side of each step. For quasi-continuous laser pumping, $T >> \tau$, and large values of $\tau$, classical laser TPA results are recovered\cite{pasquarello1,spector1} as shown in the upper inset in Fig. \ref{fig2}, where a weakly quantum correlated biphoton is assumed, $\omega_g \tau=500$. In this last case, the ETPA dependence on the biphoton entanglement amount, as given by $K>1$, is practically irrelevant,
as ETPA curves for all cases (FAC, unentangled and FC biphotons) are hard to distinguish in that plot. This behavior indicates that quantum correlated photons arriving almost simultaneously, small $\tau$ limit, should produce two-photon absorption changes near the sub-band edges, leaving unaltered the TPA values well inside the continuous bands. By contrast, when the quantum correlation window broads, large $\tau$, the classical TPA behavior is quickly recovered.
Although beyond the scope of the present work, we mention that by including a time delay between signal/idler photons a further enhancement of entanglement induced transparency should be possible as well as the exploration of virtual-state spectroscopy in a semiconductor nanostructure like a QW.

\begin{figure}[tbh]
\vspace{-1.5cm}
\hspace{-0.5cm}
\includegraphics[height=13.0cm,width=13.5 cm] {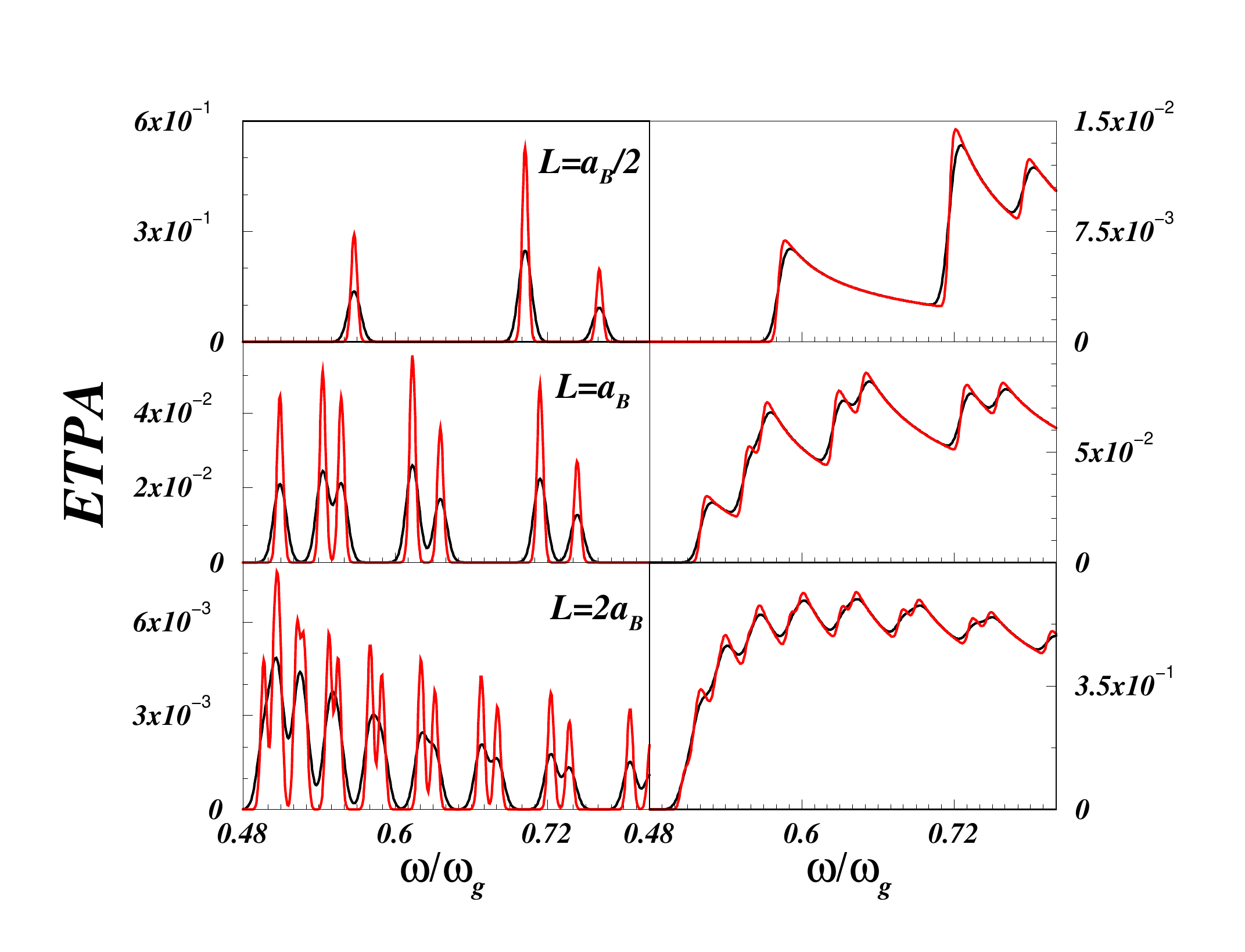}
\vspace{-1cm}
\caption{Excitonic ETPA (left column) and continuous electron-hole ETPA (right column) by QWs of thickness $L=a_B/2$ (first row), $L=a_B$ (second row) and $L=2a_B$ (third row). Red curves correspond to highly quantum correlated biphotons, $\omega_g \tau=50$ and $K=1.3$ while unentangled biphoton results, $K=1$, are represented by black curves. Notice the different vertical scales.} \label{fig4}
\end{figure}
Biphoton quantum correlations become more prominent when Coulomb-excitonic effects are taken into account. Fig. \ref{fig3} displays results where both excitonic and continuous electron-hole contributions are included. Highly quantum correlated biphotons, $\omega_g \tau=100$, produce strong excitonic peaks when the photons are frequency-anti-correlated whereas in the case of frequency-correlated photons, with the same amount of entanglement, the excitonic peaks practically vanish. This fact can be simply understood by resorting to the photon joint distribution as seen in Fig. \ref{fig1}(b) and Fig. \ref{fig1}(d), where the squeezing in both cases is evident but the spreading is higher (the maximum height is lower) in the frequency-correlated case. Although matter-radiation resonance conditions are met for different electron-hole transitions (bound and unbound) using FAC or FC biphotons, the photon joint density is larger in the former case as compared to the latter case. As a consequence, the exciton oscillator strengths are highly increased when the photons arrive almost simultaneously as can be checked by Fourier transforming the information given in Fig. \ref{fig1}(b).
\begin{figure}[tbh]
\vspace{-0cm}
\begin{center}
\includegraphics[height=9.5cm,width=12.5 cm] {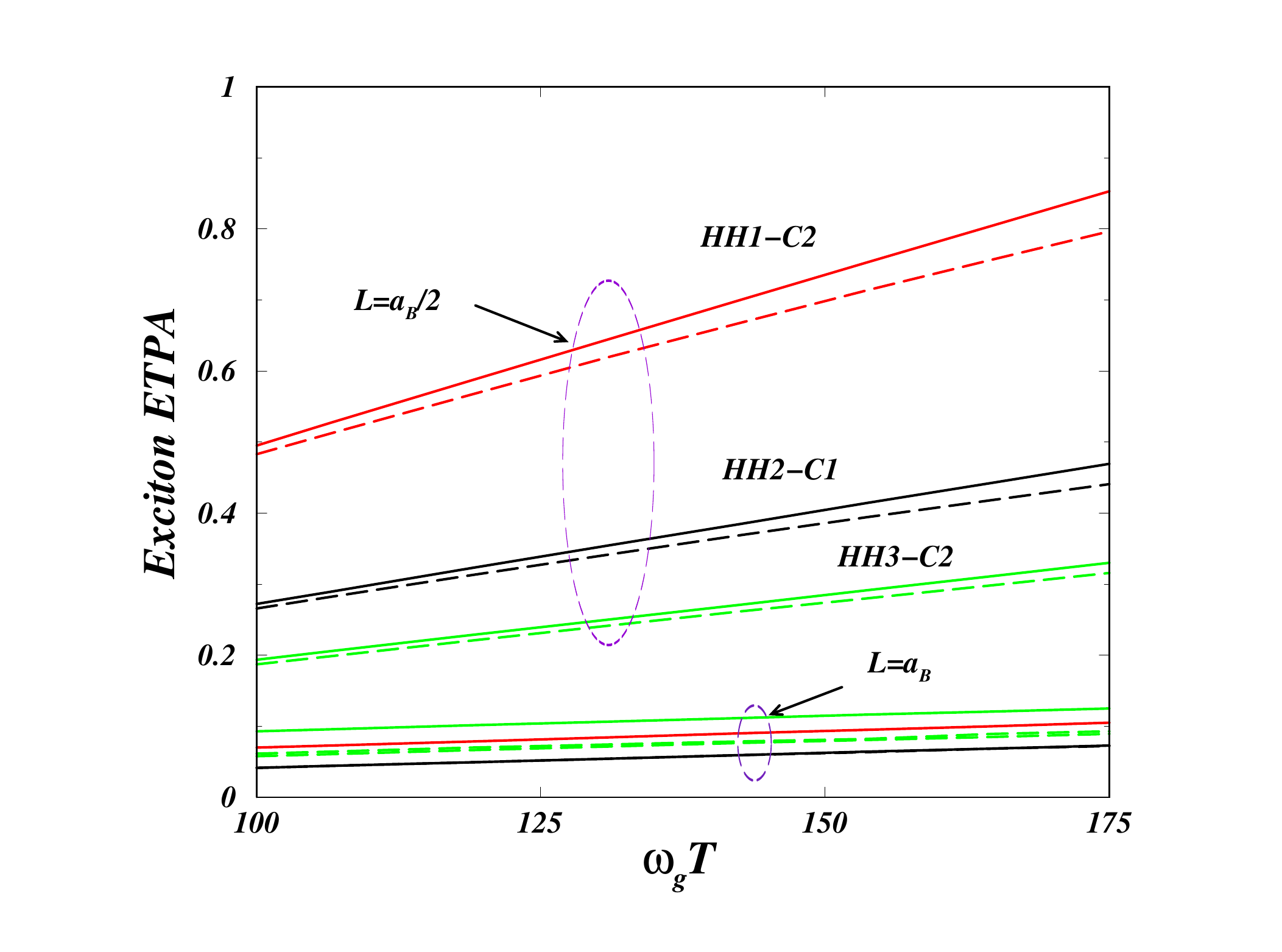}
\end{center}
\vspace{-1cm}
\caption{Excitonic ETPA as a function of the pump laser coherence time, $T$, for two QW thicknesses and different degrees of signal/idler quantum correlation times. Solid curves, $\omega_g \tau=50$. Dashed curves, $\omega_g \tau=100$. HH2-C1 exciton in black, HH1-C2 exciton in red ($X_2$ exciton state in Fig. \ref{bandstructure1})  
and HH3-C2 exciton in green.} \label{fig5}
\end{figure}

Note that these results are obtained without any assumed line-shape function for the exciton. Thus, the variation in exciton peak heights and full widths at half maximum are intrinsic and only determined by the laser pump coherence associated to time $T$, or equivalently to $K$ values, and to signal/idler temporal quantum correlations as specified by time $\tau$. The bound-exciton ETPA signal shows a highly selective and enhanced excitation efficiency under quantum-correlated illumination, somewhat similar to recently reported theoretical results for a three-level atomic system\cite{oka1}. However, the differences with the present QW case, are that selectivity and efficiency for the excitonic ETPA depend also on the specific sub-band transitions, giving higher values for the lowest-energy transitions involved. Thus, by contrast with laser induced TPA, ETPA allows the control of two-photon transitions to excitons by tailoring the frequency correlation between signal and idler photons produced by SPDC-II processes.

The interplay between carrier spatial confinement in a QW and temporal photon correlations in ETPA features is shown in Fig. \ref{fig4} for FAC biphotons (see Fig. \ref{fig1}(b)). Excitonic ETPA signals (left column) are separately displayed from continuous electron-hole ETPA signals (right column) for three QW thicknesses $L=a_B/2$, two-dimensional QW limit, $L=a_B$ and $L=2a_B$. In Fig. \ref{fig4} ETPA from highly quantum correlated biphotons with $K=1.3$ and $\omega_g \tau=50$, are depicted in red curves. In order to compare with the ETPA signal from unentangled biphotons, results for $K=1$ are also shown in black lines in Fig. \ref{fig4}, and they allow us to conclude that a strong 2D confinement enhances substantially the ETPA response from highly quantum correlated biphotons. Furthermore, Coulomb effects, enhanced by carrier confinement, also impact electron-hole or continuous ETPA results as can be seen by comparing Fig. \ref{fig2}, no Coulomb interaction, with Fig. \ref{fig4} (right column), for a narrow QW, $L=a_B/2$. Most importantly, exciton ETPA signals at exciton resonances are clearly resolved (second and third row, left column) when entangled photons are used. According to well known results, increasing the QW thickness yields to a dense packing of excitonic peaks. However, our results predict that an entangled photon source can be used to gain in spectroscopic resolution.  Additionally, the variations of the three lowest in energy excitonic ETPA peaks, as a function of the pump laser coherence time $T$, are plotted in Fig. \ref{fig5} for narrow QWs, $L=a_B/2, a_B$, and biphotons with $\omega_g \tau=50,100$ ($K > 1$). In Fig. \ref{fig5}, full ETPA signals at the resonant two-photon exciton frequencies, including the sum of the exciton and electron-hole contributions, are shown.
ETPA becomes a highly sensitive probe of photon correlations when narrow semiconductor QWs are used as two-photon absorbers. Thus, we find that one should be able to harness the non-trivial cross effects provided by carrier confinement and temporal photon quantum correlations to detect the latter ones.

Clearly, the main differences between classical TPA and ETPA results arise when strongly quantum correlated biphotons arrive to the QW absorber, i.e. small $\tau$. This is due to the fact that high temporal correlation, or equivalently time coincidence, between the photons implies the photon pair is absorbed as a whole entity and not in a sequential way of individual photon processes through an intermediate virtual state. Indeed, in the latter case classical TPA occurs. One major source of losing spectroscopic information in TPA experiments is the dephasing associated with exciton quasi-particles. However, the high sensitivity of ETPA to small quantum correlation times, $\tau$ usually in the femtosecond range, justifies that dephasing excitonic effects in the entangled-photon absorption process can be safely ignored, as it has been the case in the present work.

\section{Conclusions}
Two-photon transitions to excitons in semiconductor QWs by absorbing an entangled biphoton have been addressed. Analytical expressions for the ETPA rate including proper second-order correlation functions describing the differently quantum-correlated biphotons were obtained. We have demonstrated that optical non-linear effects at the level of a single couple of photons can be observed by playing with the correlation time window of cross-polarized photons emitted by a type-II SPDC process. In particular, we found that a high ETPA rate occurs when the sum of frequencies of the signal/idler photon pair is on resonance with the discrete QW transitions even though each photon can have a wide frequency range. This effect, specially strong for narrow QWs and highly frequency-anti-correlated biphotons, amounts to a new resource for ultrafast two-photon counting with semiconductor devices\cite{fabre2}. Furthermore, semiconductor nanostructures like QWs provide new large-bandwidth multiple-photon detectors where both discrete and continuous carrier states play important roles in discriminating the degree of photon entanglement.

It should be further mentioned that the remarkable point is that the effects here discussed does not rely on any change in the matter itself but on the quantum correlations of the exciting light. In this way a break up, or at least a weakening, of the usual optical selection rules can be tailored. We also believe that our results on excitonic ETPA can be exploited by integration with other semiconductor nanostructures, like quantum dots, in order to probe photon entanglement produced by on-chip sources. These effects can be observed in semiconductor QWs samples where the homogeneous excitonic linewidth is smaller than the entanglement induced broadening of the exciton absorption. We also suggest that cross effects from spatial carrier confinement and temporal photon quantum correlations can be exploited in various optoelectronic solid-state based entanglement set-ups providing new ways of addressing controllable quantum correlations in condensed-matter environments.

\section*{Acknowledgments}
The authors are grateful to Carlos Tejedor (UAM, Spain) and Alejandra Valencia (ICFO, Spain) for
stimulating discussions and a critical reading of the manuscript. Work supported in part by Faculty of
Sciences-Research Funds 2011 (UniAndes).

\end{document}